# An Image Steganography Scheme using Randomized Algorithm and Context-Free Grammar

**Youssef Bassil**

*LACSC – Lebanese Association for Computational Sciences*
*Registered under No. 957, 2011, Beirut, Lebanon*
*Email: youssef.bassil@lacsc.org*

**Abstract**

Currently, cryptography is in wide use as it is being exploited in various domains from data confidentiality to data integrity and message authentication. Basically, cryptography shuffles data so that they become unreadable by unauthorized parties. However, clearly visible encrypted messages, no matter how unbreakable, will arouse suspicions. A better approach would be to hide the very existence of the message using steganography. Fundamentally, steganography conceals secret data into innocent-looking mediums called carriers which can then travel from the sender to the receiver safe and unnoticed. This paper proposes a novel steganography scheme for hiding digital data into uncompressed image files using a randomized algorithm and a context-free grammar. Besides, the proposed scheme uses two mediums to deliver the secret data: a carrier image into which the secret data are hidden into random pixels, and a well-structured English text that encodes the location of the random carrier pixels. The English text is generated at runtime using a context-free grammar coupled with a lexicon of English words. The proposed scheme is stealthy, and hard to be noticed, detected, and recovered. Experiments conducted showed how the covering and the uncovering processes of the proposed scheme work. As future work, a semantic analyzer is to be developed so as to make the English text medium semantically correct, and consequently safer to be transmitted without drawing any attention.

## 1 Introduction

For decades, people have endeavored to develop pioneering techniques for secret communication. In fact, Cryptography and Steganography are two popular techniques intended to protect and safely transmit secret data. The former scrambles information so that it becomes unreadable by unauthorized parties; whereas, the latter conceals the very existence of information by embedding it into a carrier medium such as image, audio file, video file, or text file [1]. In this respect, steganography can be considered as a stealthy method for secret communication as it hides the existence of communication, so much so that no one apart from the sender and the receiver would suspect any piece of data being communicated.

Fundamentally, steganography is an information hiding technology that covers data into digital media files. Its applications are diverse, including secret communication, copyright protection, digital watermarking, and tamper proofing [2]. In practice, steganography works as follows: A message that needs to be secretly communicated is first encoded into a digital carrier file such as an image file. Then, the carrier is transmitted to the intended recipient, who upon reception, decodes it and eventually recovers the covered secret message. Obviously, the biggest advantage of steganography is that it hides the fact that a secret communication is taking place; thus, avoiding the detection of the secret message by eavesdroppers and malicious parties [3].

The strength of steganography resides in how strong the carrier medium is imperceptible and how much the covered message is difficult to be detected and uncovered by unauthorized observers. In critical situations, people known as steganalysts are hired to identify suspicious files and detect whether or not they contain secret information, and if possible, recover this information. Actually, developing a steganography algorithm that firmly conceals data in a hard-to-notice, hard-to-detect, and hard-to-recover way ensures that the secret information being communicated through certain carrier medium would pass undetected by forensics and illicit third parties.

This paper proposes a novel steganography scheme for hiding digital data into uncompressed image files using a randomized algorithm. The proposed scheme uses two mediums to deliver the secret data. The first medium is a carrier image holding the secret data inside the LSBs of its pixels which, unlike traditional LSB techniques, these pixels are selected randomly and not in sequence. The second medium is a well-structured and syntactically correct English text made up of several English sentences pointing to the location of the random carrier pixels, that is, the location of the secret data in the carrier image. In effect, the second medium is not predefined but dynamically generated during the encoding process using a mini-version context-free grammar of the English language coupled with a lexicon of English words randomly categorized in 10 categories representing the 10 digits of the decimal system. These digits are used to generate all possible location values for the carrier pixels. The proposed scheme has such advantages as being hard-to-notice, hard-to-detect, hard-to-recover, binary-based, multilingual, and mutable. All in all, they enable it to be used for versatile types of data, in total secrecy, and without getting detected or recovered, tricking stego-analysts and misleading them from the true location of the covert data.

## 2 Elements of Steganography

Algorithmically, steganography has two processes, one for covering and one for uncovering secret data. The covering process is about hiding overt data into a cover medium, also known as stego or carrier file. In contrast, the uncovering process is just the reverse; it is about extracting the covert data from the carrier file and returning them back to their original state. Fundamentally, modern digital steganography is governed by five key elements. They are as follows [4]:

1. Covert Data: Often known as the payload and refers to the overt data that need to be covertly communicated or stored. The covert data can be anything convertible to binary format, from simple text messages to executable files.
2. Carrier Medium: It is basically a file into which the covert data are concealed. The carrier medium can be any computer-readable file such as image, audio, video, or text file.
3. Stego File: Sometimes called package, it is the resulting file which has the covert data embedded into it.
4. Carrier Channel: It denotes the file type of the carrier, for instance, BMP, JPG, MP3, PDF, etc.
5. Capacity: It denotes the amount of data the carrier file can hide without being distorted.

## 3 State-of-the-Art in Image Steganography

So far, massive research work has been conducted in the development of steganography for digital images. One of the earliest techniques is the LSB technique which obscures data communication by inserting the secret data into the insignificant parts of the pixels of an image file, more particularly, into the least significant bits (LSB) [5]. The modified version of the image, which is called carrier file or stego file, is then sent to the receiver through a public channel. The foremost requirement of the LSB technique is that it should not exhibit any visual signs in the carrier image so as to not give any indications that secret data are being communicated covertly.
Basically, the LSB technique is an insertion-based image steganography method that embeds secret data into uncompressed computer image files such as BMP and TIFF. In this technique, the data to hide are first converted into a series of bytes, then into a series of smaller chunks each of which is of size *n* bits. Then, *n* LSBs of the pixels of the carrier image are replaced by each of the chunks of the original data to hide. The ultimate result of this operation is a carrier image carrying the secret data into the LSBs of its pixels. As the color values that are determined by LSBs are insignificant to the naked eye, it is hard to tell the difference between the original image and the tampered one, taking into consideration that no more than a certain number of LSBs were used to conceal the secret data; Otherwise, visual artifacts and damages would be produced in the carrier image which would in turn draw suspicions and raise attention about something unusual in the carrier image. For instance, in 24-bit True Color BMP images, using more than three LSBs per color component to hide data may result in perceptible artifacts in the carrier image [6]. As an illustration for the LSB technique, let's say that the letter H needs to be hidden into an 8-bit grayscale bitmap image. The ASCII representation for letter H is 72 in decimal or 01001000 in binary. Assuming that the letter H is divided into four chunks each of 2 bits, then four pixels are needed to totally hide the letter H. Moreover, assuming that four consecutive pixels are selected from the original image whose grayscale values are denoted by $P_1$=11011000, $P_2$=00110110, $P_3$=11001111, and $P_4$=10100011, then substituting every two LSBs in every of these four pixels by a 2-bit chunk of the letter H, would result in a new set of pixels denoted by $P_1$= 110110**01**, $P_2$=001101**00**, $P_3$=110011**10**, and $P_4$=101000**00**. Despite changing the actual grayscale values of the pixels, this has little impact on the visual appearance of the carrier image because characteristically, the Human Visual System (HVS) cannot differentiate between two images whose color values in the high frequency spectrum are marginally unalike [7].

On the other hand, other steganography techniques and algorithms for digital images have been proposed and researched both in spatial and frequency domains. They include masking and filtering [8], encrypt and scatter [9], transformation [10], and BPCS [11] techniques.

**Masking and Filtering Technique:** This technique is based on digital watermarking but instead of increasing too much the luminance of the masked area to create the digital watermark, a small increase of luminance is applied to the masked area making it unnoticeable and undetected by the naked eye. As a result, the lesser the luminance alteration, little the chance the secret message can be detected. Masking and filtering technique embeds data in significant areas of the image so that the concealed message is more integral to the carrier file.

**Encrypt and Scatter Technique:** This technique attempts to emulate what is known by White Noise Storm which is a combination of spread spectrum and frequency hopping practices. Its principle is so simple; it scatters the message to hide over an image within a random number defined by a window size and several data channels. It uses eight channels each of which represents 1 bit; and consequently, each image window can hold 1 byte of data and a set of other useless bits. These channels can perform bit permutation using rotation and swapping operations such as rotating 1 bit to the left or swapping the bit in position 3 with the bit in position 6. The niche of this approach is that even if the bits are extracted, they will look garbage unless the permutation algorithm is first discovered. Additionally, the encrypt and scatter technique employs DES encryption to cipher the message before being scattered and hidden in the carrier file.

**Transformation Technique**: This technique is often used in the lossy compression domain, for instance, with JPG digital images. In fact, JPG images use the discrete cosine transform (DCT) to perform compression. As the cosine values cannot be calculated accurately, the DCT yields to a lossy compression. The transformation-based steganography algorithms first compress the secret message to hide using DCT and then integrate it within the JPG image. That way, the secret message would be integral to the image and would be hard to be decoded unless the image is first decompressed and the location of the hidden message is recovered.

**BPCS Technique**: This technique which stands for Bit-Plane Complexity Segmentation Steganography, is based on a special characteristic of the Human Visual System (HVS). Basically, the HVS cannot perceive a too complicated visual pattern as a coherent shape. For example, on a flat homogenous wooden pavement, all floor tiles look the same. They visually just appear as a paved wooden surface, without any indication of shape. However, if someone looks closely, every collection of tiles exhibits different shapes due to the particles that make up the wooden tile. Such types of images are called vessel images. BPCS Steganography makes use of this characteristic by substituting complex regions on the bit-planes of a particular vessel image with data patterns from the secret data.

# 4 Proposed Solution

This paper proposes a novel steganography scheme for hiding any form of digital data into uncompressed digital image files in a random manner. It uses two medium to convey the secret data. An uncompressed image file acting as a carrier image holding the secret data inside the LSBs of its pixels, and an English text made up of several well-structured and syntactically correct English sentences pointing to the location of the carrier pixels, that is, the location of the secret data in the carrier image. Algorithmically, the proposed scheme is based on a randomized algorithm to randomly select the carrier pixels into which the secret data are be concealed, and on a mini-version context-free grammar of the English language to generate correct English sentences that encode the location of the random carrier pixels in the carrier image. The carrier image is processed as a 2D plane with coordinates x and y (x, y) representing pixels' locations. The employed context-free grammar uses a lexicon of English words randomly categorized in 10 categories representing the 10 digits of the decimal numeral system (0…9). These digits are used to generate all possible coordinate values for the carrier pixels.

## 4.1 The Proposed Scheme in Details

The proposed scheme is designed to work on 24-bit True Color uncompressed digital images such as BMP. Basically, the pixels of a 24-bit BMP image are each composed of three 8-bit color components, namely R, G, and B components [12]. The proposed scheme hides the secret data into the three LSBs of each of these color components; thus, the hiding capacity is equal to 9 bits out of 24 bits or 37% of the total size of the carrier image (9/24=0.375=37%). The carrier image is manipulated as a 2D plane geometric object composed of a finite set of points along with their coordinates. These coordinates effectively indicate the locations of the pixels in the carrier image itself. Figure 1 depicts a BMP image as a geometric 2D plane as regarded by the proposed scheme.

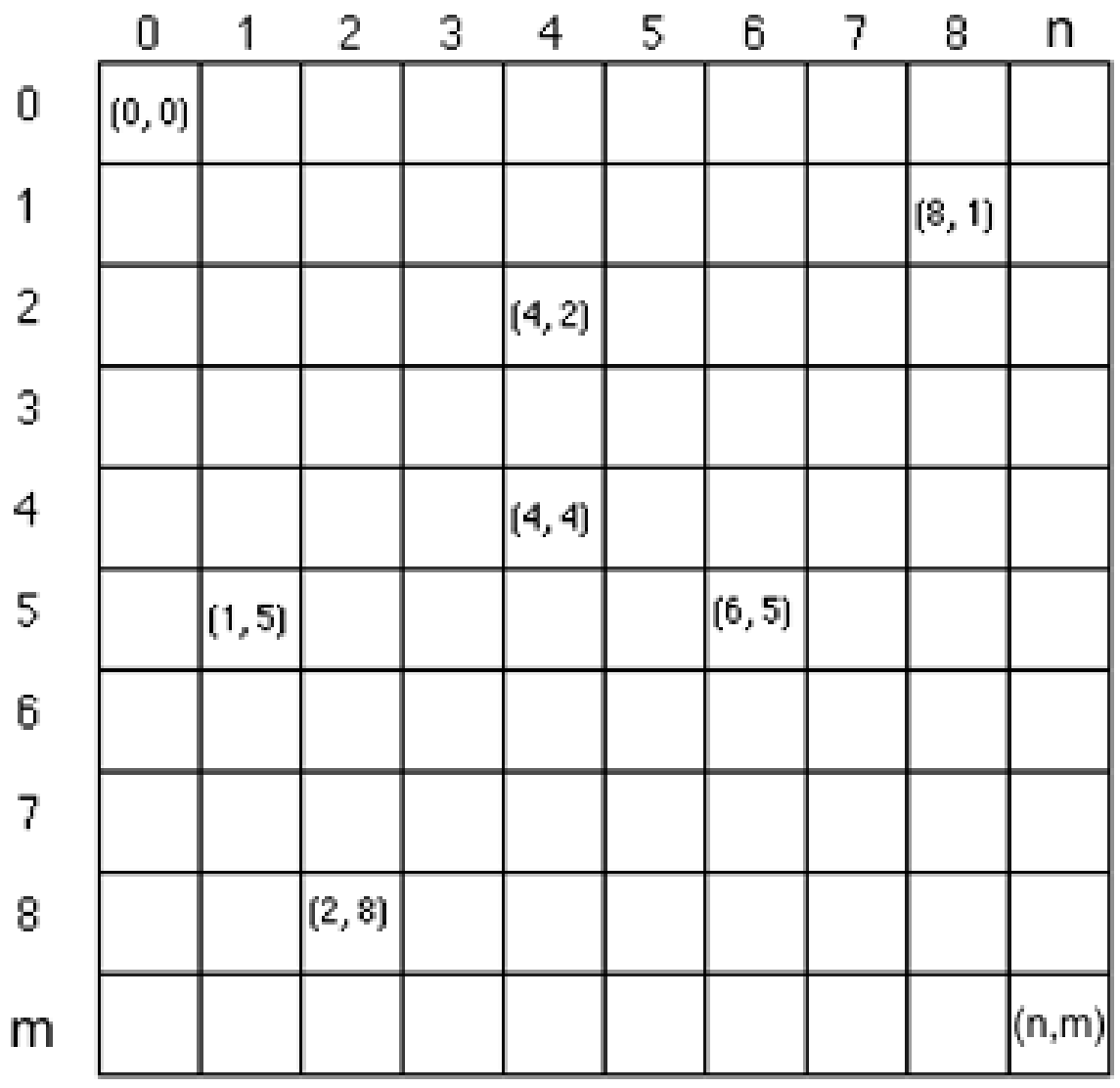

Fig. 1: Carrier Image as a 2D Plane with Coordinates

In effect, the proposed scheme does not hide the secret data sequentially into the carrier pixels; instead, it randomly selects pixels to carry in the secret data. Thus, shuffling and dispersing the secret data over the carrier image in a random manner. The actual positions of these randomly selected carrier pixels, which interchangeably are their coordinates, are mimicked by an English text composed of English sentences that are grammatically well structured. The English text is generated using a context-free grammar and a lexicon of words organized into 10 categories from category 0 till category 9. Words that constitute the English sentences are selected from these categories based on the coordinates of the carrier pixels. For instance, coordinates (019,421) are encoded by selecting a word from category 0, a word from category 1, a word from category 9, a word from category 4, a word from category 2, and a word from category 1 respectively. Actually, every coordinate (i.e. pixel location) is represented by an English sentence. As a result, multiple pixels would result in multiple sentences which eventually result in a complete English text. The context-free grammar is essential in order to generate a grammatically correct text, for instance, generating a sentence composed of a determinant, followed by a noun, followed by verb, followed by a preposition. It is worth noting that no common words exit between the categories of the lexicon as it would be later impossible to recover the real coordinates of the carrier pixels out the English text. Finally, together, the carrier image and the generated English text are sent to the receiver who has to use the same algorithm, the same grammar, and the same lexicon to recover the different locations of the carrier pixels and consequently the secret data.

## 4.2 The Context-Free Grammar and the Lexicon

A context-free grammar or CFG is a mathematical system for modeling the structure of languages such as natural languages like English, French and Arabic, or computer programming languages like C++, Java, and C# [13]. Its formalism was originally set by Chomsky [14] and Backus [15], independently of each other. The one of Backus is known as the Backus-Naur Form or BNF for short [16]. In essence, a context-free grammar consists of a set of rules known as production rules that specify how symbols and words of a language can be arranged and grouped together. An example of a rule would specify that in the English language a verb phrase "VP" must always start with a verb then followed by a noun phrase "NP". Another rule would state that a noun phrase "NP" can start with a proper noun "ProperNoun" or a determinant "Det". Following is a sample CFG for a hypothetical language called L, that is a subset of the English language.

*NP → Det Nominal*
*NP → ProperNoun*
*Nominal → Noun | Nominal Noun*
*Det → an*
*Det → the*
*Noun → apple*

The symbols that are used in a CFG are divided into two classes: Terminals that correspond to unbreakable words in the language such as "apple", "the", and "an"; and non-terminals which are variables that can be replaced by terminals and other non-terminal variables to derive and produce sentences of the language.

Terminals are usually provided by a lexicon of words; whereas, the non-terminals and the production rules are part of the parsing algorithm. Parsing is the process of deriving sentences for the language using the production rules of the CFG and the lexicon [17]. Parsing can also be used to confirm that the structure of a sentence complies with the CFG of the language.
The proposed steganography scheme uses a mini-version CFG of the English language and a lexicon of predefined words to generate correct English sentences that can encode the coordinates of the carrier pixels. Figure 2 outlines the production rules of the CFG used by the proposed scheme.

$$
\begin{aligned}
&S \rightarrow NP\ VP\\
&S \rightarrow VP\\
&NP \rightarrow Pronoun\\
&NP \rightarrow Proper\text{-}Noun\\
&NP \rightarrow Det\ Nominal\\
&Nominal \rightarrow Noun\\
&Nominal \rightarrow Nominal\ Noun\\
&Nominal \rightarrow Nominal\ PP\\
&VP \rightarrow Verb\\
&VP \rightarrow Verb\ NP\\
&VP \rightarrow Verb\ NP\ PP\\
&VP \rightarrow Verb\ PP\\
&VP \rightarrow VP\ PP\\
&PP \rightarrow Preposition\ NP
\end{aligned}
$$

Fig. 2: CFG of the Proposed Scheme

On the other hand, the lexicon of the CFG which is outlined in Table 1 is organized into 10 categories each of which contains a set of terminal words that belong to the English language. It is worth mentioning that no common words exist between the categories of the lexicon as it would be later impossible to recover the real coordinates of the carrier pixels out of the English text.

Table 1: Lexicon of the Proposed CFG

<table>
<tr><th colspan="2">Category 0</th><th>Category 1</th></tr>
<tr><td colspan="2">Det → this | that | …<br>Pronoun → I | they | us | …<br>Preposition → from | across | about | …<br>Noun → door | car | memory | …<br>Verb → play | eat | walk | …<br>Proper-Noun → California | John | Intel | …</td><td>Det → those | an | …<br>Pronoun → he | she | me | …<br>Preposition → on | above | through | …<br>Noun → tree | school | board | …<br>Verb → study | dance | climb | …<br>Proper-Noun → Texas | NBA | …</td></tr>
<tr><th>Category 2</th><th>…</th><th>Category 9</th></tr>
<tr><td>Det → a | these | …<br>Pronoun → you | it | their | …<br>Preposition → to | towards | along | …<br>Noun → girl | university | roof | …<br>Verb → sleep | like | enroll | …<br>Proper-Noun → Ohio | George | Mike | …</td><td></td><td>Det → the | …<br>Pronoun → we | ours | his | …<br>Preposition → among | at | before | …<br>Noun → floor | book | pen | …<br>Verb → move | walk | write | …<br>Proper-Noun → Harvard | Tony | …</td></tr>
</table>

Every word in the above categories uniquely encodes the category to which it belongs. For instance, the word "this" encodes the digit "0" exclusively because it belongs to category 0. The word "Harvard" encodes the digit "9" exclusively because it belongs to category 9, and so forth. It is not mandatory, in practice, that the lexicon of Table 1 is used exactly as is; the communicating parties can compile their own lexicon and use it mutually for the covering as well as the uncovering operation.

## 4.3 The Proposed Algorithm

Below is the list of steps executed by the proposed algorithm to hide a piece of secret data into an image file.

1. The secret data are preprocessed so that they become suitable for storage inside the carrier image.
   a. The secret data, no matter their types and formats – whether text, documents, or executables, are converted into a binary form resulting into a string of bits denoted by D={$b_0, b_1, b_2, b_3, \ldots b_{n-1}$} where $b_i$ is a single bit composing the secret data and $n$ is the total number of bits.

- **b.** The string of bits D is organized into chunks of 3 bits, such as D=Chunks={ $C_0[b_0, b_1, b_2]$, $C_1[b_3, b_4, b_5]$, $C_2[b_6, b_7, b_8]$,…$C_{m-1}[b_{n-3}, b_{n-2}, b_{n-1}]$ }, where $C_j$ is a particular 3-bit chunk, *m* is the total number of chunks, and *n* is the total number of bits making up the secret data.

2. A set of carrier pixels is randomly selected from the carrier image in which every random selected pixel is denoted by $P_t(x_t, y_t)$ where *x* and *y* are the coordinates of pixel P, and *t* is the index of P.
3. The chunks of the secret data that were created in step 1.b are embedded into the randomly selected pixel of step 2.
   - **a.** As the carrier image is a 24-bit colored image, every pixel would have three color components R, G, B, each of length 8 bits. For this reason, every chunk $C_j$ is stored in the three LSBs of each of the color components of every selected carrier pixel such as $P_t$={ *fiveMSBs*($R_t$) + $C_j$ ; *fiveMSBs*($G_t$) + $C_{j+1}$ ; *fiveMSBs*($B_t$) + $C_{j+2}$ }, where P is a randomly selected pixel, *t* is the index of P, and R, G, and B are the three Red-Green-Blue color components of pixel $P_t$. Furthermore, *fiveMSBs*(component) is a function that returns the original five most significant bits of the color component "component". The "+" operator concatenates the original five MSBs of the color component with the 3 bits of a particular chunk, making the total number of bits in a given color component equals to 8 bits. In effect, the first 5 bits are the original five MSBs of the color component and the three LSBs are a particular chunk from the secret data.
4. The coordinates of the randomly selected pixels of step 2 are encoded into English sentences each of which is denoted by $S_k$.
   - **a.** Every single digit of the coordinates *x* and *y* such as $x=\{d_0,d_1,d_2,d_{r-1}\}$ and $y=\{d_0,d_1,d_2,d_{r-1}\}$ is mapped into a category number of the lexicon. For instance, P(29, 01) is encoded using the CFG of Figure 2 and the lexicon of Table 1 as "a book from school". In that, "a" is a determinant from category 2, "book" is a noun from category 9, "from" is a preposition from category 0, and "school" is a noun from category 1.
   - **b.** Step 4.a is repeated for all selected pixels. Eventually, the number of all generated English sentences would be equal to the number of all randomly selected pixels.
5. The final output comprises two mediums. The first one is the carrier image which houses the secret data into its randomly selected pixels such as IMG=$\{P_0,P_1,P_2,P_{t-1}\}$, where P is a carrier pixel and *t* is the total number of carrier pixels; while, the second one is an English text made up of grammatically correct English sentences such as T=$\{S_0,S_1,S_2,S_{t-1}\}$, where S is an English sentence and *t* is the total number of English sentences which is also equal to the total number of carrier pixels. Both, the carrier image and the English text are to be sent to the receiver, not necessarily at the same time, however, they should be both present when the receiver is uncovering the secret data.

# 5 Experimentations & Results

An example is illustrated in this section to demonstrate how the proposed steganography scheme works. It involves all the steps required to cover a secret text message into a 24-bit BMP carrier image file. The carrier image is sampled at 800x600 resolution, and the secret data to hide is a text message denoted by D="kill joe".

1. The secret message is preprocessed and converted into a binary form.
   - **a.** The secret message in ASCII format can be denoted as $D_{ASCII}$={ k=1101011 i=1101001 l=1101100 l=1101100 space=0100000 j=1101010 o=1101111 e=1100101 }
   - **b.** Chunks of 3 bits are generated out of D and are denoted by Chunks={ $C_0$[110], $C_1$ [101], $C_2$[111], $C_3$[010], $C_4$[011], $C_5$[101], $C_6$[100], $C_7$[110], $C_8$[110], $C_9$[001], $C_{10}$[000], $C_{11}$[001], $C_{12}$[101], $C_{13}$[010], $C_{14}$[110], $C_{15}$[111], $C_{16}$[111], $C_{17}$[001], $C_{18}$[01]. The total number of chunks is 19; and therefore, 19 color components are required to store these chunks. Since every pixel has three color components, this requires 19/3 = 6.3 = 7 pixels to store the different chunks of the secret message D.
2. Seven pixels are randomly selected. Their coordinates are respectively: $P_0$(206, 318), $P_1$(407, 192), $P_2$(321, 129), $P_3$(709, 015), $P_4$(501, 000), $P_5$(712, 200), and $P_6$(309, 108).
   The color values of these seven pixels are as follows: $P_0$(R=00101111 ; G=00111111 ; B=10101010) , $P_1$(R=11101001 ; G=10110011 ; B=00111011) , $P_2$(R=10100000 ; G=00001111 ; B=00101000) , $P_3$(R=11101111 ; G=11111111 ; B=10000000) , $P_4$(R=11101001 ; G=00110000 ; B=10101111) , $P_5$(R=10101010 ; G=00000000 ; B=11111111) , $P_6$(R=11110010 ; G=11111111 ; B=10000011).
3. The chunks obtained in step 1.b are stored into the three LSBs of the color components of every pixel of step 2. The results are as follows: $P_0$(R=00101110 ; G=00111101 ; B=10101111) ; $P_1$(R=11101010 ; G=10110011 ; B=00111101) ; $P_2$(R=10100100 ; G=00001110 ; B=00101110) ; $P_3$(R=11101001 ; G=11111000 ; B=10000001) ; $P_4$(R=11101101 ; G=00110010 ; B=10101110) ; $P_5$(R=10101111 ; G=00000111 ; B=11111001) ; $P_6$(R=11110010 ; G=11111111 ; B=10000011). The red bits are chunks

of the secret message that replaced the original three LSBs of the color components of the selected carrier pixels. The hiding capacity of the algorithm is equal to 9 bits out of 24 bits or 37% of the total size of the carrier image (9/24=0.375=37%). Figure 3 depicts the original carrier image before and after hiding the secret message D inside its pixels.

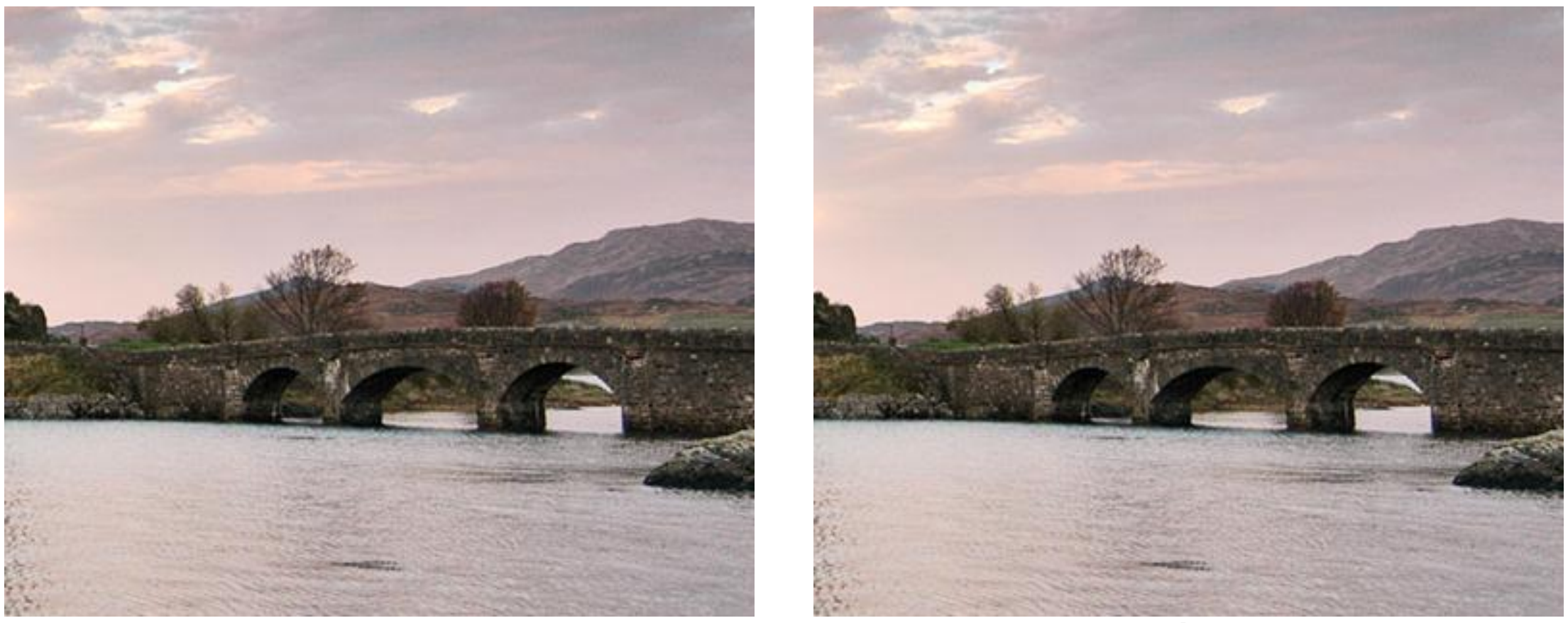

a b

Fig. 3: Before (a) and After (b) the covering process

**4.** The coordinates of the randomly selected pixels are encoded into English sentences using the CFG and the lexicon of Figure 2 and Table 1. Since there are seven pixels, there should be seven English sentences. Table 2 outlines the generated English sentences each denoted by S along with the coordinates of the corresponding pixels.

Table 2: The Generated English Sentences

| | Category/ Word | Category/ Word | Category/ Word | Category/ Word | Category/ Word | Category/ Word |
|---|---|---|---|---|---|---|
| $P_0$(206,318) | 2 | 0 | 6 | 3 | 1 | 8 |
| $S_0$ | the | boy | went | to | high | school |
| | | | | | | |
| $P_1$(407,192) | 4 | 0 | 7 | 1 | 9 | 2 |
| $S_1$ | eat | the | apple | in | the | kitchen |
| | | | | | | |
| $P_2$(321,129) | 3 | 2 | 1 | 1 | 2 | 9 |
| $S_2$ | John | write | circle | on | the | wall |
| | | | | | | |
| $P_3$(709,015) | 7 | 0 | 9 | 0 | 1 | 5 |
| $S_3$ | dance | in | the | night | with | mike |
| | | | | | | |
| $P_4$(501,000) | 5 | 0 | 1 | 0 | 0 | 0 |
| $S_4$ | a | computer | shipped | to | Atlanta | city |
| | | | | | | |
| $P_5$(712,200) | 7 | 1 | 2 | 2 | 0 | 0 |
| $S_5$ | George | does | homework | on | the | desk |
| | | | | | | |
| $P_6$(309,108) | 3 | 0 | 9 | 1 | 0 | 8 |
| $S_6$ | these | guys | dive | in | the | ocean |

**5.** The final output is the carrier BMP image now carrying the secret data, and the English text denoted by T={ “the boy went to high school”, “eat the apple in the kitchen”, “john write circle on the wall”, “dance in the night with mike”, “a computer shipped to Atlanta city”, “George does homework on the desk”, “these guys dive in the ocean” }. It is worth noting that the algorithm does not guarantee that the generated English sentences are all semantically correct as it does not implement a semantic analyzer. It is the job of the user to compile the lexicon with the appropriate words that can semantically work with each other.

As for the uncovering process, it is the reverse of the above process. First, the second medium which is the English text T is decomposed into sentences such as T={$S_0$,$S_1$,$S_2$,$S_{t-1}$}. Then, each sentence is decomposed into words. Then, every word is mapped to a digit pertaining to the category in the lexicon it belongs to. At this point, coordinates of the carrier pixels are generated and are used to extract the secret data from the three LSBs of the pixels they point to. As a result, a string of bit is formulated which is then converted into ASCII text, revealing the original secret message D, mainly “kill joe”.

# 6 Advantages of the Proposed Scheme

The proposed steganography scheme has several advantages over other existing schemes. They are listed below.

The first advantage is that the proposed scheme is hard-to-notice, a fact that is promoted by using only three LSBs in every color component, in addition to a well-structured English text to convey secret data. It is by using only three LSBs to hide data, ensures that no visual artifacts are to be produced in the carrier image. Likewise, a grammatically correct text lowers suspicions and totally obscures the fact that a secret communication is taking place.

The second advantage is that the proposed scheme is hard-to-detect, a fact that is promoted by selecting the carrier pixels in a random manner. Using random pixels shuffles the secret data and scatters them all over the carrier image. That way, no one apart from the holder of the English text, the lexicon, and the decoding algorithm can know how to extract the secret data out the carrier image.

The third advantage is that the proposed scheme is hard-to-recover, a fact that is promoted by using two mediums to deliver the secret data, mainly the carrier image and the English text. As a result, the scheme is less susceptible to stego-analysis attacks as third parties often assume that the secret data are hidden in one medium and not between two mediums that complement each other. Using the proposed scheme, the sender can first send one of the mediums, and then later on, send the other one.

The fourth advantage is the capacity of data that can be hidden. As the proposed scheme can hide secret data into the three LSBs of each of the color components of every pixel, the hiding capacity is then equal to 9 bits out of 24 bits or 37% of the total size of the carrier image (9/24=0.375=37%). As a result, a 1MB BMP carrier image can carry 1*37%=0.37MB=378KB, which is around one third of the total image size. All that while retaining the visual quality of the carrier image.

The fifth advantage is that the proposed scheme is binary-based which enables it to operate on text messages as well as images, audio files, video files, applications, word documents, HTML and XML documents, spreadsheets, PDF documents, and any sort of data that can be converted into a stream of bits. This makes the scheme so generic and suitable for a wide range of applications.

The sixth advantage is that the proposed scheme is multilingual, in that, it can be tailored to support languages other than the English language by just providing the appropriate CFG and lexicon for that language. In fact, almost any natural language in the world has a grammar and a set of words proper to build a lexicon. As a result, the scheme can be used by many users regardless of their native language.

The seventh advantage is that the proposed scheme is mutable, in such a way that it is not bound to a specific lexicon and new lexicons can be compiled and loaded into the algorithm. As a result, new lexicons can derive new sentences for the same coordinate values, fooling off stego-analysts and impeding them from achieving their attacks.

# 7 Conclusions & Future Work

This paper proposed a random steganography scheme for hiding digital information into digital image files. Unlike, major steganography methods which use only one medium to transmit the secret information, the proposed scheme uses two mediums, mainly a carrier image embedding the actual secret data and a well-structured text encoding the random locations of the carrier pixels. The scheme proved to be advantageous over other methods. It is hard-to-notice as it preserves the quality of the carrier image; it is hard-to-detect as it randomly selects carrier pixels; it is hard-to-recover as it uses two mediums that complement each other to convey the secret information; it is binary-based as it can hide any type of digital information and not necessarily text; it is multilingual as it can generate the text medium in different languages depending on the CFG and lexicon used; and it is mutable as it can generate different text mediums for the same pixels’ coordinates by just changing the content of the lexicon. All in all, the proposed scheme outstandingly obscures data in such an anonymous way that the secret communication would pass undetected by forensics and passive eavesdroppers.

As future work, the proposed scheme can be improved by adding to it a semantic analyzer so as to generate semantically structured English sentences for the text medium, which consequently, would be more innocent-looking and safer to be transmitted without drawing any attention.

## Acknowledgements

This research was funded by the Lebanese Association for Computational Sciences (LACSC), Beirut, Lebanon, under the “Stealthy Steganography Research Project – SSRP2012”.